\documentstyle[preprint,aps]{revtex}
\begin{document}

\title{
Exact Critical Exponents for Pseudo-Particles in the Kondo Problem
}
\author{S. Fujimoto$^1$, N. Kawakami$^2$, and S.-K. Yang$^3$}
\address{
$^1$Department of Physics, Faculty of Science,
Kyoto University, Kyoto 606, Japan\\
$^2$Department of Material and Life Science,
and Department of Applied Physics, \\
Osaka University, Suita, Osaka 565, Japan\\
$^3$Institute of Physics, University of Tsukuba, Ibaraki 305, Japan
}
\maketitle
\begin{abstract}
Exact critical exponents of the Green functions for
pseudo-fermions and slave bosons in the
SU($N$) Anderson model with $U\rightarrow\infty$
are obtained by using the Bethe ansatz solution and boundary
conformal field theory. They are evaluated exactly
for mixed valence systems and Kondo systems
with crystalline fields. The results agree with the prediction of
Menge and M\"uller-Hartmann, which coincide
with those of the X-ray problem.
Some implication of our results in one-dimensional chiral systems
is also discussed.
\end{abstract}
\pacs{PACS numbers: }
Some ten years ago, the slave boson method was proposed as a
tool to deal with the strong correlation problem
for interacting electron systems\cite{slave1,slave2}.
The essence of the method is that in the presence of the
strong Coulomb interaction, the original Fock space of
electrons can be mapped to that of pseudo-particles, so-called the
slave boson, $b$, and the pseudo-fermion,
$f_{m}$ ($m=1,2, \cdots, N$), under the constraint for
the conservation of particles,
$b^{\dagger}b+\sum_{m}f^{\dagger}_{m}f_{m}=1$.
More recently, Menge and M\"uller-Hartmann studied
low-energy dynamical properties of these
pseudo-particles for the impurity Anderson
model with infinite Coulomb interaction
by using the bosonization method\cite{menge}.
In the simplest case with degeneracy $N=1$, they obtained
the critical exponents of the Green functions which coincide
with those found in the X-ray problem\cite{noz}.
They also predicted the formula of
the exponents for general SU($N$) cases,
and the result for the $N=2$ case was recently confirmed
numerically by the renormalization group method\cite{costi}.
It is interesting  to study  the critical behavior
of pseudo-particles {\it exactly}, because we can directly see the
infrared catastrophe inherent in the Kondo problem
through their asymptotic behavior.

In this paper, we obtain the exact critical
exponents for pseudo-particles of the
$U \rightarrow \infty$ SU($N$) Anderson
model by combining the Bethe ansatz solution\cite{sch}
with the finite-size scaling in
boundary conformal field theory\cite{cardy}.
The form of the exponents we shall derive is in agreement with that
obtained by Menge and M\"uller-Hartmann\cite{menge}. We evaluate them
explicitly for mixed-valence and Kondo regimes.
In the previous paper, we pointed out that
one can read anomalous exponents related to the X-ray
edge singularity from the finite-size spectrum of the Anderson model,
by using boundary conformal field theory\cite{fky}
(see also ref.\cite{al}).
We apply  this idea to the present problem,
obtaining the exact critical exponents.

The Hamiltonian we consider is
the $U \rightarrow \infty$ SU($N$) Anderson model,
\begin{eqnarray}
H&=&\sum_{m=1}^N \int{\rm d}x c^{\dagger}_{m}(x)\Big( -i
\frac{\partial}{\partial x} \Big) c_{m}(x)
+\epsilon_{f}\sum_{m=1}^N f_m^{\dag}f_m \nonumber \\
& &+V\sum_{m=1}^N \int{\rm d}x\delta (x)
\Big( f_m^{\dag} b c_{m}(x)+c_{m}^{\dagger}(x)
b^{\dag} f_m \Big),
\label{eqn:hamsun}
\end{eqnarray}
with the constraint $b^{\dagger}b+\sum_{m}f^{\dagger}_{m}f_{m}=1$,
where the impurity electron states have $N$-fold degeneracy.
The Hamiltonian (\ref{eqn:hamsun}) has been reduced to
the one-dimensional one by using partial-wave
representation, and the spectrum for
conduction electrons has been linearized
around the Fermi energy.
The pseudo-particle operators $b$ and $ f_m$ $(m=1,2,\cdots,N)$
represent the unoccupied and single
occupied states at the impurity site,
and the double occupancy is strictly forbidden due to
the infinite Coulomb repulsion.

The exact finite-size spectrum of the model\cite{fky}
was obtained from the Bethe-ansatz solution\cite{sch}.
We briefly summarize the results necessary for the present
discussions. The effective two-body $S$-matrix
for host electrons which
satisfies the Yang-Baxter equation is
\begin{equation}
S(k_i-k_j)=\frac{k_i-k_j-iV^2 P_{\alpha \beta}}{k_i-k_j-iV^2}.
\end{equation}
where $P_{\alpha\beta}$ is a permutation operator
of two coordinates $x_\alpha$ and $x_\beta$, and
the effect of the impurity is incorporated through the
phase shift due to the impurity,
$\exp(2i\phi(k_j))$ with
 $\phi(k)= {\rm tan^{-1}} (2(k-\epsilon_f)/V^2)$.
Under periodic boundary conditions, one can then diagonalize
the many-body $S$-matrix, and
obtain the Bethe ansatz equations\cite{sch}.
The finite-size excitation spectrum computed by the standard
technique\cite{dw,ize,de,suzu} is written in terms of
the $N \times N$ matrix\cite{fky},
\begin{equation}
\frac{1}{L}E_1=
\frac{2\pi v}{L}\frac{1}{2}\Delta {\bf M}^{T}
{\cal C}_{f}\Delta {\bf M}
-\frac{\pi v}{L}N
\biggl(\frac{\delta_F}{\pi}\biggr)^2, \label{eqn:fss1}
\end{equation}
where  $\Delta M^{(l)} \equiv
\Delta M_h^{(l)}-\frac{\delta_F}{\pi}(N-l)$ for $1\leq l\leq N-1$,
and $\Delta  M^{(0)}=\Delta N_h-N\delta_F/\pi$
with $\delta_F$ being the phase shift at the Fermi level.
Here $\Delta N_h$ is the number of charge excitations,
$\Delta M_h^{(l)}$'s are quantum numbers related to
spin degrees of freedom, and the $N \times N$ matrix
${\cal C}_{f}$ is given by
\begin{equation}
{\cal C}_{f}=
\left(
\matrix {1     & -1      & \null  &
                 \smash{\lower1.7ex\hbox{\LARGE 0}} \cr
        -1     &  2  & \ddots & \null   \cr
        \null  & \ddots  &  \ddots     & -1      \cr
        \smash{\hbox{\LARGE 0}}   & \null   &  -1    & 2  \cr}
\right) .
\end{equation}
One can check that the last term in eq.(\ref{eqn:fss1}),
which has been evaluated from the excited states,
is equal to the shift of the ground-state energy due to the presence
of the impurity\cite{shift}. Therefore the
increment of the ground state energy
cancels the last term of eq.(\ref{eqn:fss1}),
which is thus irrelevant for
the discussions of critical exponents and will be dropped
in the following.

We wish to generalize the above results to
more general cases with magnetic fields and crystalline
fields. In these cases, the phase shifts $\delta_l$ depend on
the total angular momentum, $l$. The first-order
energy corrections in $1/L$ due to external fields
read
\begin{equation}
E^{(1)}=\sum_{m=0}^{N-1}\frac{\delta_{l}}{\pi}\frac{\Delta N_l}{L},
\end{equation}
where $\Delta N_l$ is the number of added conduction
electrons with total angular momentum $l$.
This term together with the $1/L$-corrections to the
finite-size spectrum of host
electrons gives the total finite-size spectrum, which
is given by eq.(\ref{eqn:fss1}) with the quantum numbers replaced by,
$\Delta M^{(0)}\rightarrow\Delta N_h-\sum_{l=0}^{N-1}\delta_l/\pi$,
and $\Delta M^{(l)}\rightarrow
\Delta M_h^{(l)}-\sum_{i=l}^{N-1}\delta_i/\pi$ for $1\leq l\leq N-1$.
These formulae are most general and are
useful to discuss the case with a crystalline field.

We are now ready to study the long-time behavior of the
Green functions for pseudo-particles;
$\langle f^{\dagger}_{m}(t)f_{m}(0)\rangle\sim t^{-\alpha_f}$, and
$\langle b^{\dagger}(t)b(0)\rangle\sim t^{-\alpha_b}$.
Applying the finite-size scaling we can determine
the critical exponents of correlation functions.
As discussed previously \cite{fky},
we can neglect the phase shifts,
if we are concerned with canonical exponents characterizing the
local Fermi liquid, because the phase-shift effect is equivalent to imposing
twisted boundary conditions, and such effect can be incorporated into
the redefinition of the charge quantum number\cite{fky}.
In order to derive critical
exponents for pseudo-particles, however,
{\it we must regard the fractional number of $f$-electrons,
$n_l=\delta_l/\pi$, as quantum numbers}, and thus the
phase shifts play an essential role to determine the
critical exponents.  For instance, to obtain
the Green function of pseudo-fermions,
we set the quantum numbers as $\Delta N_h=1$ and
$\Delta M_h^{(l)}=0$. We then read off
the corresponding critical exponent,
\begin{equation}
\alpha_f=1-\frac{2\delta_F}{\pi}+N\biggl(\frac{\delta_F}{\pi}
\biggr)^2,
\end{equation}
in the absence of crystalline fields and magnetic fields.
The exponent for slave-boson Green function, $\alpha_b$,
is obtained in a similar manner. Since the
slave boson expresses a vacancy, it carries neither charge nor spin.
Thus by taking $\Delta N_h=\Delta M_h^{(l)}=0$, one has
\begin{equation}
\alpha_b=N\biggl(\frac{\delta_F}{\pi}\biggr)^2.
\end{equation}
The formulae for $\alpha_f$ and $\alpha_b$ agree with those predicted by
Menge and M\"uller-Hartmann by means of
bosonization\cite{menge}, and take the same form
as those in the X-ray problem:
the exponent of pseudo-fermion is equal to the X-ray
absorption exponent, and that of slave boson
to the X-ray photoemission exponent.
We have evaluated the critical exponents
$\alpha_{f,b}$ exactly for the SU($N$) model with total angular
momentum $J=1/2, 3/2, 5/2, 7/2$ ($N=2J+1$),
which have been shown as a function of
the renormalized energy level of $f$-electrons, $\epsilon^{*}_f$,
in figs.1 and 2. Note that $J=5/2$ case ($J=7/2$ case)
corresponds to Ce (Yb) impurities in a metal.

We next discuss the crystal-field effects on
the critical exponents. The effect of crystalline fields are
particularly important for {\it Kondo systems} where $\epsilon_f$
lies far below the Fermi level. As such an example,
we consider the SU(6) Coqblin-Schrieffer model
($\epsilon_f \rightarrow -\infty$ limit) with
cubic symmetry, which is considered as an
appropriate  model for Ce impurities in a metal\cite{sch,ko}.
In this model, the charge fluctuation is completely suppressed,
and six-fold degenerate states split into a
$\Gamma_7$ doublet and a $\Gamma_8$ quartet in a
cubic crystalline field.
For the case of the $\Gamma_7$-ground state,
we set the phase shifts as $\delta_0=\delta_1\equiv \delta_{\Gamma_7}
\neq \delta_2=\delta_3=\delta_4=\delta_5\equiv \delta_{\Gamma_8}$.
Thus from the finite-size spectrum with a crystalline field,
we obtain the exponent of pseudo-fermions of
the $\Gamma_7$ doublet by taking $\Delta N_h=1$, $\Delta M_h^{(l)}=0$,
\begin{equation}
\alpha_f^7=\frac{5}{6}-\frac{4}{3}(n_{\Gamma_7}-n_{\Gamma_8})
+\frac{4}{3}(n_{\Gamma_7}-n_{\Gamma_8})^2,
\end{equation}
where $n_{\Gamma_7}=\delta_{\Gamma_7}/\pi$, and
 $n_{\Gamma_8}=\delta_{\Gamma_8}/\pi$.
In the case of the  $\Gamma_8$ ground state,
we set $\delta_0=\delta_1=\delta_2=\delta_3\equiv \delta_{\Gamma_8}
\neq \delta_4=\delta_5\equiv \delta_{\Gamma_7}$.
Then taking  $\Delta N_h=1$, $\Delta M_h^{(l)}=0$,
we obtain the exponent of pseudo-fermion of the $\Gamma_8$ quartet,
\begin{equation}
\alpha_f^8=\frac{5}{6}+\frac{2}{3}(n_{\Gamma_7}-n_{\Gamma_8})
+\frac{4}{3}(n_{\Gamma_7}-n_{\Gamma_8})^2.
\end{equation}
On the other hand, for the case of  slave boson, by taking
$\Delta N_h=0$, $\Delta M_h^{(l)}=0$, one gets
\begin{equation}
\alpha_b=\frac{1}{6}+\frac{4}{3}(n_{\Gamma_7}-n_{\Gamma_8})^2.
\end{equation}
We computed the critical exponents
$\alpha_f^{7,8}$ and $\alpha_b$ by the exact solution, and
plotted them in fig.3 as a function of the crystal-field splitting,
$\Delta\epsilon=\epsilon_{\Gamma_8}-\epsilon_{\Gamma_7}$.

Finally, we comment on some implication of the above results in
one-dimensional (1D) chiral systems.
The Fermi-edge singularity problem in 1D electron systems
has attracted current interest\cite{lut1,lut2,lut3}.
If host electrons move only in one direction and
the backward scattering due to the impurity is irrelevant,
the results obtained above are applicable
to 1D systems with a slight
modification. Such a situation may be realized
 in the edge state of the quantum Hall effect\cite{edge}.
Thus the following results may be the case for
the Fermi-edge singularity in the fractional
quantum Hall effect. In 1D Luttinger liquids, the Luttinger
parameter (charge correlation exponent),
$K_{\rho}$, appears in the finite-size spectrum of the charge sector.
Thus the spectrum is given by eq.(\ref{eqn:fss1}) with ${\cal C}_f$
replaced by
\begin{equation}
{\cal C}_{f}=
\left(
\matrix {\frac{1}{NK_{\rho}}+\frac{N-1}{N}     & -1      & \null  &
                 \smash{\lower1.7ex\hbox{\LARGE 0}} \cr
        -1     & 2  & \ddots & \null   \cr
        \null  & \ddots  &  \ddots     & -1      \cr
     \smash{\hbox{\LARGE 0}}   & \null   &  -1    & 2  \cr}
\right) .
\end{equation}
Then the critical exponent for the X-ray absorption in this system
is
\begin{equation}
\alpha_f=\frac{1}{NK_{\rho}}\biggl(1-\frac{N\delta}{\pi}\biggr)^2
+\frac{N-1}{N},
\end{equation}
and that for the photoemission is
\begin{equation}
\alpha_b=\frac{N}{K_{\rho}}\biggl(\frac{\delta}{\pi}\biggr)^2.
\end{equation}
Note that for the edge state of the fractional quantum Hall effect
with filling $\nu=N/(Nm+1)$ ($m$ even),
$K_{\rho}$ is solely determined by the  filling factor $\nu$ as
$K_{\rho}=\nu/N$. We expect such anomalous exponents may be
observed in the X-ray problem
in the edge state of the fractional quantum Hall effect.

The authors are grateful to A. W. W. Ludwig for valuable discussions.
This work was partly supported by a Grant-in-Aid from the Ministry
of Education, Science and Culture, Japan.


\begin{figure}
\caption{
Critical exponent for pseudo-fermion
$\alpha_f$ as a function of  $\epsilon_f^{*}/2\Delta$ for
the SU($2J+1$) Anderson model. The resonance width is $\Delta=V^2/2$.}
\label{fig.1}
\end{figure}
\begin{figure}
\caption{Critical exponent for
slave boson $\alpha_b$ as a function of $\epsilon_f^{*}/2\Delta$.}
\label{fig.2}
\end{figure}
\begin{figure}
\caption{Critical exponents
$\alpha_f$ for $\Gamma_7$ (broken line) and $\Gamma_8$ (solid line),
and $\alpha_b$ as a function of
the crystal-field splitting
 $\Delta\epsilon/T_0$ (
 $\Delta\epsilon=\epsilon_{\Gamma_7}-\epsilon_{\Gamma_8}$)
for the SU(6) Coqblin-Schrieffer model with cubic symmetry.
Here $T_0$ is the Kondo temperature for $\Delta \epsilon =0$.}
\label{fig.3}
\end{figure}



\begin{references}
\bibitem{slave1} S. E. Barnes, J. Phys. {\bf F6} (1976) 1375;
{\bf 7} (1977) 2637.

\bibitem{slave2} P. Coleman, Phys. Rev. {\bf B29} (1984) 3035.

\bibitem{menge} Menge and M\"uller-Hartmann, Z. Phys. {\bf B73} (1988)
225.

\bibitem{noz} P. Nozi\`eres and de Dominicis, Phys. Pev. {\bf 178} (1969)
1097.

\bibitem{costi} T. A. Costi, P. Schmetteckert, and J. Kroha, and P. W\"olfle,
Phys. Rev. Lett. {\bf 73} (1994) 1275.

\bibitem{sch} P. Schlottmann, Phys. Rev. Lett. {\bf 51} (1983) 1697;
Phys. Rep. {\bf 187} (1989) 1.

\bibitem{cardy} J. L. Cardy, Nucl. Phys. {\bf B240} (1984) 514;
{\bf B324} (1989) 581; J. L. Cardy and D. C. Lewellen, Phys. Lett. {\bf B259}
(1991) 274.

\bibitem{fky} S. Fujimoto, N. Kawakami, and S. -K. Yang, Phys. Rev.
{\bf B50} (1994) 1046.

\bibitem{al} I. Affleck and A. W. W. Ludwig, J. Phys. {\bf A27} (1994)
5375.

\bibitem{dw} H. J. de Vega and F. Woynarovich, Nucl. Phys. {\bf B251},
439 (1985); F. Woynarovich, J. Phys. {\bf A22}, 4243 (1989).

\bibitem{ize} A. Izergin, V. E. Korepin and  N. Yu Reshetikhin,
J. Phys. {\bf A22} (1989)  2615.

\bibitem{de} H. J. de Vega, J. Phys. {\bf A21} (1988) L1089.

\bibitem{suzu} J. Suzuki, J. Phys. {\bf A21} (1988) L1175.

\bibitem{shift}
Though it is not straightforward to
compute the shift of the ground state energy
for the model (\ref{eqn:hamsun})
because of the linearized spectrum of host electrons,
one can deduce the correct shift, for example, by
taking the ordinary dispersion $E=k^2$.

\bibitem{ko} N. Kawakami and A. Okiji, J. Phys. Soc. Jpn. {\bf 54}
(1985) 685.

\bibitem{lut1} T. Ogawa, A. Furusaki, and N. Nagaosa, Phys. Rev. Lett.
{\bf 68} (1992) 3638.

\bibitem{lut2} C. L. Kane, K. A. Matveev and L. I. Glazman,
Phys. Rev. {\bf B49} (1994) 2253.

\bibitem{lut3} N. V. Prokof'ev, Phys. Rev. {\bf B49} (1994) 2148.

\bibitem{edge} X. G. Wen, Int. J. Mod. Phys. {\bf B6} (1992) 1711.

\end{references}
\end{document}